\documentclass{ctr_summer}

\usepackage{ctrfont}
\usepackage{natbib}
\usepackage{undertilde}
\usepackage{color}
\usepackage[labelsep=period]{caption}
\usepackage{verbatim}
\usepackage[dvips]{graphicx}
\usepackage{psfrag}
\usepackage{epsfig}
\usepackage{amsmath,rotating}
\let\square\relax
\usepackage{amssymb}
\usepackage{dashrule}
\usepackage{undertilde}
\usepackage{subcaption}
\usepackage{url}

\usepackage{color}
\newcommand{\norm}[1]{\lVert #1 \rVert}

\newcommand{\Es}{E^{\rm s}}
\newcommand{\Eu}{E^{\rm u}}
\newcommand{\Xs}{X^{\rm s}}
\newcommand{\Xu}{X^{\rm u}}
\title{Toward computing sensitivities of average quantities in turbulent flows}

\shorttitle{Chaotic sensitivity computation}

\author{
N. Chandramoorthy\footnote{Department of Mechanical Engineering, 
Massachusetts Institute of Technology}, 
Z. N. Wang\footnote[3]{\label{camb}Department of Engineering, University of Cambridge, UK},
Q. Wang\footnote[2]{Department of Aeronautics and Astronautics,
Massachusetts Institute of Technology} \and P. Tucker\footnotemark[3]}
\shortauthor{Chandramoorthy et~al.}

\begin{document}

%% Setting the first page number.  Please leave as it is, at "1".
\setcounter{page}{1}

\maketitle

Chaotic dynamical systems such as turbulent flows are characterized by an exponential divergence of infinitesimal 
perturbations to 
initial conditions. Therefore, conventional adjoint/tangent sensitivity analysis methods that are successful with RANS 
simulations fail in the case of chaotic LES/DNS. In this work, we discuss the limitations of current approaches, including ensemble-based and shadowing-based sensitivity methods, that were proposed as alternatives to conventional
sensitivity analysis. We propose a new alternative, called the space-split sensitivity (S3) algorithm, that is computationally
efficient and addresses these limitations. In this work, the derivation of the S3 algorithm is presented in the special case where the 
system converges to a stationary distribution that can be expressed 
with a probability density function everywhere in phase-space. Numerical examples of low-dimensional chaotic 
maps are discussed where S3 computation shows good agreement with finite-difference results, indicating potential 
for the development of the method in more generality.
\\

\hrule

%% SECTION NAMES:  
%% Should be lower-case except for proper nouns and abbreviations
%% Should not end with a period
%=========================================================================
\section{Introduction}
%=========================================================================

%OPEN AS MANY SECTIONS AS REQUIRED (EXAMPLE BELOW)
%=========================================================================
%=========================================================================

Conventional tangent/adjoint sensitivity analysis methods have been
extensively applied to RANS simulations for uncertainty quantification, mesh adaptation and gradient-based 
multidisciplinary design optimization applications \citep{patrick, samareh}. 
Many modern applications require computing sensitivities in DNS/LES; examples include buffet prediction in high-maneuverability 
aircraft, modern turbomachinery design and jet engine and
airframe noise control. The methodologies for sensitivity analysis
in these high-fidelity simulations must be more sophisticated than the conventional tangent/adjoint approaches
since the latter are known to produce meaningless sensitivities of statistically stationary quantities under
chaotic dynamics \citep{angxiu,qiqi}. 

In the next section, we discuss the deficiencies of current alternative methods, including 
ensemble-based and shadowing-based methods. In particular, ensemble-based methods are prohibitively
expensive owing to the inherent instability and associated high uncertainty of tangent, adjoint and 
finite-difference methods in chaotic systems. Shadowing-based methods do not always compute 
the correct sensitivities since they are based on determining stable, shadowing perturbations that are not guaranteed to carry the correct average information about the flow.
The purpose of this work is to develop a new methodology, called the space-split sensitivity
(S3) algorithm, that addresses these deficiencies to 
produce a provably convergent and computationally efficient means to compute sensitivities of 
statistical averages to input parameters. 

The strategy used by S3 to circumvent the ill-conditioning
of tangent and adjoint equations deviates from that of both ensemble-based and shadowing-based approaches.
While ensemble sensitivity suffers because of the unbounded variance of the
unstable contribution to the overall sensitivity, S3 splits the contributions and performs a
finite-sample averaging only in order to obtain the stable
contribution. The unstable contribution is manipulated through integration-by-parts and with ergodic properties of the fluid flow to yield an algorithm that does not
use unstable tangent solutions. Since both parts of the sensitivity are computed through
sampling on generic flow trajectories, the problem of the computed sensitivities corresponding
to unrepresentative trajectories, which shadowing-based methods are vulnerable to, is averted. 
In Section \ref{sec:derivation}, we elucidate these key ideas of stable-unstable splitting and of the 
modification of the unstable contribution. We describe the algorithm derived under simplifying mathematical assumptions in Section \ref{sec:algorithm} and demonstrate the algorithm on low-dimensional numerical 
examples in Section \ref{sec:results}.

\section{Current methods and their limitations}
\label{sec:review}
Consider a chaotic map parameterized by a set of parameters $s$,
\begin{align}
	u_{i+1} = \varphi^s (u_i), \: i = 0, 1, 2 \cdots , u_i \in \mathbb{R}^n.
\end{align}
In a fluid simulation, the state vector $u_i$ consists of all the unknowns
at the timestep $i$, such as the density, 
velocity components and pressure at all the grid points, and in this case, 
$n = 5 \times $ the number of grid points. 
The transformation $\varphi^s$ is the Navier-Stokes
solver that advances the state by one timestep, with examples of $s$ being geometric parameters of the domain or
inlet conditions and so on. The fluid state $u_i$
can be written as a function of the initial state $u_0$ as
$u_i = \varphi^s_i(u_0)$, where the subscript $i$ in $\varphi_i^s$
refers to solving for $i$ timesteps; that is, $\varphi_i^s = \underbrace{\varphi^s \circ 
\cdots \circ \varphi^s}_{i\; {\rm times}}$. We also use the notation $\varphi_{-i}^s$ to denote the transformation $(\varphi^s)^{-1}$ composed with itself $i$ times to indicate solving backward in time by $i$ timesteps; that is, $\varphi_{-i} (u_i) = u_0$.

We are concerned with statistically stationary fluid systems 
where the state vector 
has achieved a stationary probability distribution $\mu^s$ in phase-space. 
The superscript
$s$ in $\mu^s$ indicates that the distribution depends on the input parameters. We are 
interested in determining the sensitivity of the statistical average, with respect to the 
distribution $\mu^s$, of 
a scalar objective function $J$ denoted by $\langle J \rangle := \int J \; d\mu^s$, to $s$. Examples of objective 
functions include lift and drag over wings and pressure losses in turbine wakes.
Under the assumption of ergodicity, the statistical average of a bounded function $J$
is also observed as an infinite time average along almost every flow trajectory. That is,
if $J \circ \varphi_i^s(u_0)$ is the instantaneous value of $J$ at timestep $i$ starting 
at $u_0$, then $\lim_{N \to \infty} (1/N) \sum_{i=0}^{N-1} J \circ \varphi_i^s (u_0) = \langle J \rangle$ for almost every initial condition $u_0$. 
This infinite time average, called the ergodic average, is the more natural form of $\langle J\rangle$ from the simulation 
standpoint, since it can 
be obtained by measurements of $J$ along trajectories. The ergodic average up to a large
$N$ is used in practice to approximate the 
ensemble average $\langle J \rangle$. 

\subsection{Ill-conditioned conventional tangent and adjoint methods}
\label{sec:conventional_tangent_adjoint}

When the notation $v_i(u_0) := (d\varphi_i^s/ds)(u_0)$ is used, the 
familiar tangent equation derived by means of a linear approximation of the transformation $\varphi_i^s$ around the reference value 
of $s$ is given by,
\begin{align}
	\notag
	v_{i+1}(u_0) &= D\varphi^s (u_i)  v_{i}(u_0) + \frac{\partial \varphi^s}{\partial s} (u_i), \; i = 0, 1, \cdots \\
	v_0(u_0) &= \frac{d u_0}{ds} = 0 \in \mathbb{R}^n,
	\label{eqn:tangent}
\end{align}
where $(D\varphi^s)(u_i)$ is the Jacobian matrix evaluated
at $u_i$ [$D f$ denotes differentiation of a function $f$ with respect 
to a phase point and $(Df)(u)$ refers to the value of the derivative 
at the point $u$]. In a conventional tangent sensitivity computation, we simply 
apply the chain rule to calculate the instantaneous sensitivity of the 
scalar field $J \circ \varphi_{i}^s$ as $d_s (J \circ \varphi_{i}^s) = (D J)(u_i)\: v_i $. 
In any chaotic system, $\norm{v_i(u_0)} \sim {\cal O}(\exp(\lambda i)),$ 
$\lambda > 0$ (the largest among the so-called Lyapunov exponents), 
for almost every $u_0$; therefore, the instantaneous 
sensitivity $d_s(J \circ \varphi_{i}^s)$ 
also grows in norm exponentially with $i$. Since $\langle J \rangle$ is equal 
to its infinite time ergodic average, one may naturally try to compute
$d_s \langle J \rangle$ by using the instantaneous sensitivities obtained
with the tangent vectors as $\lim_{N \to \infty} (1/N) \sum_{i=0}^{N-1} 
 (D J)(u_i)\: v_i$. But the latter quantity is unbounded, whereas the correct
 sensitivity is a finite quantity, thereby rendering the sensitivities computed
 from the tangent equation meaningless for large $i$. Since the adjoint 
 equation when solved backward in time also has exponentially diverging
 solutions, sensitivities 
 computed by using the adjoint method are also unbounded for large $i$.

\subsection{Ensemble sensitivity analysis and its computational expense}
\label{sec:ensemble_sensitivity}

The Lea-Allen-Haine ensemble sensitivity method \citep{eyink} suggests a simple work-around to 
the exponentially diverging sensitivities computed by the conventional tangent/adjoint methods. The work-around is 
to truncate the values of $i$ at a finite $N$ that represents an intermediate timescale 
on the same order as $1/\lambda$ and instead introduce phase-space averaging over a finite
sample of independent trajectories. The rigorous justification for this approximation is given by a statistical response formula
due to \cite{ruelle-maps}, which describes the sensitivity we want to compute as a summation,
where each summand is a phase-space average
\begin{align}
	\label{eqn:ruelle}
		\frac{d \langle J \rangle}{d s} &=
		\sum_{i=0}^\infty \int D(J \circ \varphi^s_{i+1}) \; \frac{\partial \varphi^s }{ \partial s } \; d \mu^s.
\end{align}
The formula states that although the integrand is unbounded as $i\to \infty$
for almost every trajectory, the integral is bounded at every $i$ 
because of cancellations
that occur on averaging over 
phase-space. The ensemble sensitivity computed with Eq. \ref{eqn:ensemble_sensitivity}
is an approximation of Ruelle's formula. For a detailed analysis of the 
ensemble sensitivity methods and fluid flow examples 
that show that Ruelle's formula is not practically computable, see \cite{nisha}.

\subsection{Inconsistency of the non-intrusive least squares shadowing method}
\label{sec:nilss}
An alternative method for computing $d_s \langle J \rangle$ is the non-intrusive
least squares shadowing (NILSS) method (see \cite{patrick,angxiu,angxiu-nilsas,qiqi} for 
details). The method computes a shadowing perturbation that remains 
bounded in a long time window under the tangent dynamics. 
The sensitivity computed by using the shadowing tangent solution is 
not guaranteed to be an unbiased estimate of the true sensitivity. 
This is because while ergodic sums converge for almost
every trajectory as noted earlier, the measure zero subset of the attractor on which they do not converge
is nontrivial. Some well-known examples of such subsets include unstable periodic orbits that form a dense subset 
of the attractor (\cite{upo}). We therefore seek an alternative that does not rely on computations along a single 
trajectory that is not guaranteed to be typical.
\section{The space-split sensitivity algorithm derivation}
\label{sec:derivation}
As noted in Section \ref{sec:review}, 
Ruelle's response
formula computed with a Monte-Carlo summation 
has unbounded variance in general since the tangent vector field
$\partial_s \varphi^s$ has components that are 
unstable under time evolution. However, in the special case
that a tangent vector is stable under time evolution, the variance
of the ensemble sensitivity estimate does not increase with time.
Therefore, we can compute the sensitivity
using the conventional tangent method. This leads to the first
step of the S3 algorithm: to split the stable and unstable 
components of the tangent vector field. That is, we  
first convert Ruelle's response formula to a tangent space-split
form as
\begin{align}
	\label{eqn:ruelle_split}
		\frac{d \langle J \rangle}{d s} &=
		\sum_{i=0}^\infty \int D(J \circ \varphi^s_{i}) \; \Xs \; d \mu^s + \\ 
		\notag
		&\sum_{i=0}^\infty \int D(J \circ \varphi^s_{i}) \; \Xu \; d \mu^s, 
\end{align}
where we use the notation $X(u_0) := (\partial \varphi^s/\partial s) \circ (\varphi^s)^{-1} (u_0)$ and
$X = \Xs + \Xu $. The mathematical 
characterization of dynamical systems, wherein we can achieve
this splitting, is hyperbolicity. In a hyperbolic dynamical system,
the tangent space at every point in phase-space can be decomposed into 
stable and unstable subspaces, denoted $\Es$ and $\Eu$ respectively, such 
that the norm of a tangent vector in $\Es$ decays exponentially while the norm of 
a tangent vector in $\Eu$ grows exponentially in time. 
Under the hyperbolicity assumption, the vector field $X$ can 
be split (note that this is not an orthogonal decomposition but a direct sum decomposition) into the vector fields 
$\Xs$ and $\Xu$ such that 
at each $u_0$, $\Xs(u_0) \in \Es(u_0)$ and 
$\Xu(u_0) \in \Eu(u_0)$. That is, there exist 
$\lambda \in \left(0,1\right), c > 0$ such that for all $i > 0$,
\begin{align}
	\label{eqn:hyperbolicity}
	\norm{D\varphi^s_{-i} \Xu(u_0)} \leq c \lambda^i \norm{\Xu(u_0)} \\
\notag	\norm{D\varphi^s_i \Xs(u_0)} \leq c \lambda^i \norm{\Xs(u_0)}.
\end{align}

The first term on the right-hand side of Eq.(\ref{eqn:ruelle_split}) is the stable contribution to the sensitivity
and the second term is the unstable contribution. 
The stable
contribution is given approximately by the following summation,
\begin{align}
	\label{eqn:stable_contribution}
	\frac{d \langle J \rangle}{d s}_{\rm stable}  &\approx (1/N) \sum_{i=0}^{N-1} 
		DJ \circ \varphi^s_{i}(u_0) \; v_i (u_0).	
\end{align}
In the equation above, $v_i$ refers to the stable tangent solution, 
that is, the conventional tangent solution with the 
unstable components subtracted from the source term 
at every timestep. Although 
we have computed an ergodic summation along a finite trajectory,
the variance of the estimate does not 
increase with $i$. Thus, the stable contribution can be computed accurately simply
by solving the tangent equation along a long trajectory.

\subsection{Regularization of the unstable contribution}

In order to compute $(d \langle J\rangle /d s)_{\rm unstable}$, we seek 
a computable transformation of Ruelle's formula. In the unstable contribution 
as expressed in Eq. (\ref{eqn:ruelle_split}), 
the phase-space average, computed as a Monte-Carlo estimate, 
has unbounded variance. On the other hand, the integral representing the phase-space average is itself 
bounded for all $i$. Therefore, we first perform integration-by-parts on the integral
since that has a regularization effect on the unbounded integrand. For simplicity, 
we now suppose that
the probability distribution is smooth
in the sense that we can write $\mu^s(u_0) = \rho^s(u_0) \; du_0$, where 
$\rho^s$ is an invariant probability density function with a compact support. 
In uniformly hyperbolic systems with a compact attractor, the sufficient
condition for the existence of $\rho^s$ is that the Jacobian determinant 
${\rm det} D\varphi^s_i$ is bounded for all $i$ \citep{katok}. Using the identity ${\rm div}(f X) = f {\rm div}(X) + 
Df \cdot X$ from vector calculus, where $f$ is a smooth 
scalar field and $X$ is a vector field, we write the
the $i$th integral as
\begin{align}
	\notag
	\int D(J \circ \varphi^s_{i}) \cdot \Xu \; \rho^s \: dx &= 
	\int {\rm div}(
        J \circ \varphi^s_{i} \:
	 \rho^s \: \Xu) \; d x \\ 
	\label{eqn:vector_identity}
	&- \int J \circ \varphi^s_{i} \:
	{\rm div} (\rho^s \Xu)  \; d x.   
\end{align}
Since $X^{\rm u}$ is a tangent vector field, the first term above would be 
zero because it would reduce to the integral of a flux function that is 
everywhere zero on a subset of 
$\mathbb{R}^{n-1}$, on application of Stokes theorem. 
Thus, the above equation, which amounts to performing an integration-by-parts, 
does achieve a regularization because now, the integrand of the second term above 
is bounded for all $i$, although it is nonsmooth for large $i$. 
\subsection{Expressing as ergodic averages}
We have expressed the unstable contribution as a phase-space average 
of a bounded quantity. However, there is no natural way to compute the phase-space
integral [the second term in Eq. (\ref{eqn:vector_identity})] since 
we can compute only ergodic averages along trajectories and this integral 
is not of the form of an ergodic average. In order to achieve a computable
form, we first use the vector identity again to obtain the two terms
\begin{align}
	\label{eqn:vector_identity_again}
	\frac{d\langle J\rangle}{d s}_{\rm unstable} 
		&= - \sum_{i=0}^\infty \Big( \int J \circ \varphi^s_{i} \:
	\frac{D\rho^s}{\rho^s} \cdot \Xu \; \rho^s  \; d x  \\
	\notag
	&+ \int J \circ \varphi^s_{i} \:
	{\rm div} \Xu  \rho^s \; d x \Big). 
\end{align}

The second term in Eq. (\ref{eqn:vector_identity_again}) above
is equivalent to an ergodic average at almost every $u_0$ and 
can be approximately evaluated on a trajectory of finite length. Moreover,
the fact that the probability density $\rho^s$ is unknown does 
not pose a problem to the computation of the second term.
The first term, however, needs to be manipulated in order 
to be expressed as an ergodic average of a function that (a) is bounded
for all $i$ and (b) can be computed similarly to the second
term, without the knowledge of $\rho^s$. The second term 
satisfies both conditions and the problem now reduces 
to computing the first term.

\subsection{Treatment of the unknown $D\rho^s/\rho^s$}
\label{sec:psis}
We seek a method to compute the first term in Eq. (\ref{eqn:vector_identity_again}) even though $D\rho^s$ is unknown. For this we use the 
time invariance under the transformation of the stationary density $\rho^s$. 
Consider the 
$i$th summand in the first term. Using the measure-preservation property of $\varphi^s$, we obtain

\begin{align}
	\label{eqn:mpp}
	\int J \circ \varphi^s_{i} \:
	\Xu \cdot \frac{D \rho^s}{\rho^s} \: \rho^s \; d x &= 
	\int J \circ \varphi^s_{i+1} \:
	\Big(\Xu \cdot \frac{D \rho^s}{\rho^s}\Big)\circ \varphi^s \: \rho^s \circ \varphi^s \; d x.
\end{align}

Using the invariance of $\rho^s$ (or expressing the fact that $\rho^s$ is an 
eigenfunction of the Frobenius-Perron transfer operator), we have,
\begin{align}
	\label{eqn:frobeniusPerron}
	\rho^s \circ \varphi^s = \frac{\rho^s}{{\rm det}(D\varphi^s)}.
\end{align}
Thus, when $i=0$, 
\begin{align}
	\label{eqn:mpp_first_term}
	\int J \:
	\Xu \cdot \frac{D \rho^s}{\rho^s} \: \rho^s \; d x &= 
	\int J \circ \varphi^s \:
	\Big(\Xu \cdot \frac{D \rho^s}{\rho^s}\Big)\circ \varphi^s \:  \frac{\rho^s}{{\rm det}(D\varphi^s)} \; d x.
\end{align}
Our intention is to compute Eq. (\ref{eqn:vector_identity_again})
in which $D\rho^s/\rho^s$ is unknown.
Since $\Xu(u_0) \in E^{\rm u}(u_0)$, replacing 
$D\rho^s/\rho^s$ with its orthogonal projection on 
$ E^{\rm u}(u_0)$, $\Pi_{E^{\rm u}} (D\rho^s/\rho^s)$, does not
change the integral in Eq. (\ref{eqn:vector_identity_again}). The same argument holds for 
all $i$ in Eq. (\ref{eqn:vector_identity_again}). 
For ease of notation and further derivation, let us define 
\begin{align}
\label{eqn:psis_definition}
\psi^s := \Pi_{E^{\rm u}} (D\rho^s/\rho^s). 
\end{align}

Now, differentiating Eq. (\ref{eqn:frobeniusPerron}) with respect to
phase points and using this derivative to define
the pullback of $\psi^s$ through $\varphi^s$ (which 
one can also interpret as the action of the Koopman 
operator, $K$, on $\psi^s$),
\begin{align}
	\label{eqn:rho_derivative_with_psis}
	K\psi^s := \psi^s\circ \varphi^s  
	&=  (D\varphi^s)^{-1} \psi^s 
	- \frac{(D\varphi^s)^{-1}}{ {\rm det}(D\varphi^s)} D {\rm det}(D\varphi^s). 
\end{align}
In general, we can write the iterate of $\psi^s$ under $\varphi^s$ 
as a recursive equation by applying $\varphi^s$ to
Eq. (\ref{eqn:rho_derivative_with_psis}),
\begin{align}
	\label{eqn:koopman_tangent}
	K_j \psi^s := \psi^s \circ \varphi^s_j
	&=  ((D\varphi^s)^{-1} \circ \varphi^s_{j-1}) K_{j-1} \psi^s
	- Y^s \circ \varphi^s_{j-1} \\
	\label{eqn:koopman_tangent_explicit}
		&= \Big(\prod_{k=1}^{j-1}(D\varphi^s)^{-1} \circ \varphi^s_k\Big) \psi^s - 
		\sum_{k=0}^{j-1} \Big( \prod_{l=k+1}^{j-1} (D \varphi^s)^{-1} \circ \varphi^s_l \Big) Y^s \circ \varphi^s_k,
\end{align} where we have used 
\begin{align}
\label{eqn:Ys}
	Y^s := \Pi_{E^{\rm u}}\frac{(D\varphi^s)^{-1}}{ {\rm det}(D\varphi^s)} D {\rm det}(D\varphi^s). 
\end{align}

In Eq. (\ref{eqn:koopman_tangent}), consistent with our notation $\varphi_j^s$,
$K_j$ refers to the $j$-time composition of $K$, $K_j := \underbrace{ 
K \circ \cdots \circ K}_{j\;  { \rm  times}}$. Let us call 
Eq. (\ref{eqn:koopman_tangent}), a linear 
equation for the evolution of $\psi^s$, the Koopman tangent equation.
Now, if the 
Koopman tangent equation were solved for $\psi^s$ with the source term 
$Y^s$ in the unstable subspace at each timestep, the norm of the iterates
would decrease with $j$ exponentially at the rate of $\lambda$. This is because any vector in the
unstable subspace decreases in norm under the action of the inverse of the Jacobian, 
$(D\varphi^s)^{-1}$ [see Eq. (\ref{eqn:hyperbolicity})]. Following this argument,  $\psi^s \circ \varphi^s_j$ reduces 
to the second term of Eq. (\ref{eqn:koopman_tangent_explicit}) for large $j$, since the first term goes to 0.
Thus, $\psi^s (u_0)$ can be computed
by solving the Koopman tangent equation starting with $\psi^s \circ \varphi^s_{-j}(u_0) = 0$ for large $j$.

This completes the list of requirements for computing the unstable contribution.
The vector field $\psi^s$ obtained above can then be substituted into  
Eq. \ref{eqn:vector_identity_again} to compute the first term as an ergodic average,
just as we sought. It is worth noting that the ergodic average to be computed 
is in the form of a 
time correlation between $J$ and $\psi^s\cdot \Xu + {\rm div} \Xu$. We approximately compute the time correlation function over a time series with a finite number
of terms, that is, up to $i \leq L$ in Eq. (\ref{eqn:vector_identity_again})
for some $L$ and the accuracy of the approximation depends on the rate of decay 
of correlations in the system. For each $i$, the ergodic average is computed with
a single trajectory of finite length, say, $N$. Since every $i$-th 
summand requires an ergodic average to be computed, this na\"ive way leads to 
${\cal O}(L N)$ computations. Below, we present an efficient algorithm that reuses computations and has a complexity of ${\cal O}(L + N)$ to compute 
Eq. (\ref{eqn:vector_identity_again}). 
\vspace{-3.5cm}

\section{The space-split sensitivity algorithm description}
\label{sec:algorithm}
\begin{enumerate}[1.]
	\item[(a)] Solve for a primal trajectory 
	$u_i = \varphi^s_i(u_0), i = 0, 1 \cdots, K $, up to a large $K$. Set 
	 $(d\langle J\rangle /ds)_{\rm stable} =  (d\langle J\rangle /ds)_{\rm unstable} = 0,$ 
		$v_0 = \psi^s_0 = 0 \in \mathbb{R}^n$. Then, follow 
		the next steps for each $i = 0, 1 \cdots , K$. 

	\item[(b)] Compute an orthonormal basis for the unstable subspace 
		$E^{\rm u}(u_i)$ and call
	it $q^j_i$, $j = 1,\cdots, m$, where $m$ is the total number of unstable 
		modes or positive Lyapunov exponents. The procedure to obtain
		such a basis involves solving $m$ homogeneous tangent equations 
		[that is, Eq. (\ref{eqn:tangent}) with a non-zero
		initial condition and a zero source term] and orthogonalization 
		using QR decomposition at each timestep. This is similar to the
		algorithm \cite{ginelli} used in the computation of 
		covariant Lyapunov vectors. $m$ can be determined by 
		applying QR to a random orthonormal basis of an arbitrary 
		dimension $< N$. $m + 1$ is the  
		minimum dimension of the basis required so that 
		the last column of the Q matrix 
		corresponds to a stable tangent vector that decays with time.

	\item[(c)] In the same procedure, use the homogeneous 
	adjoint equation in order to obtain a basis for the 
		adjoint unstable subspace (the subspace of the dual 
		of the tangent space that consists of vectors that grow
		exponentially in time under the homogeneous adjoint equation) 
		that
		is orthogonal to $E^{\rm s}(u_i)$ and hence denoted 
		$E^{\rm s^\perp}(u_i)$. 
	Let us call the orthonormal
	basis vectors $p^j_i$, $j = 1, \cdots, m$. 

\item[(d)] At each $i$, obtain the decomposition $X_i = X^{\rm u}_i + X^{\rm s}_i$ as follows. Write
	$X^{\rm u}_i = \sum_{k=1}^m a_i^k q_i^k$ and solve for the unknown coefficients $a_i^k$ by using the orthogonality of $X^{\rm s}_i$ to $E^{\rm s^\perp}(u_i)$. That is,	solve for $a_i^k$ in $(X_i - \sum_{k=1}^m a_i^k q_i^k ) \cdot p^j_i  =  0$. Upon obtaining $a_i^k$, $\Xu_i$ is computed and $\Xs_i = X_i - \Xu_i$.
\item[(e)]  Solve the tangent Eq. (\ref{eqn:tangent}) using the source term $X^{\rm s}_i$. Obtain $v_{i}$. 

\item[(f)] Solve the Koopman tangent equation with source term $Y^s_{i-1}$ to obtain $\psi^s_{i}$.
	Note that the solution $\psi^s_{i}$ becomes more accurate with $i$, as explained 
		in Section \ref{sec:psis}, although we arbitrarily set $\psi^s_0 = 0$. 
	\item[(g)] For large enough $i \geq M$ and setting $M' = K - M + 1$, use $\psi^s_i$ and $v_{i}$ to update the unstable and stable contributions,
		respectively, as follows [using Eq. \ref{eqn:vector_identity_again} and \ref{eqn:stable_contribution}],
		\begin{align}
			\frac{d\langle J\rangle}{ds}_{\rm unstable}
			&= \frac{d\langle J\rangle}{ds}_{\rm unstable} 
			- \frac{1}{M'-i}\sum_{j=i}^{K} J_j \Big( \:
			\psi^s_i \cdot \Xu_i  
			+
			{\rm div} \Xu_i \Big)  \\
			\frac{d \langle J \rangle}{ds}_{\rm stable} &= 
		\frac{d \langle J \rangle}{ds}_{\rm stable} + 
			\dfrac{1}{M'} DJ \circ \varphi_{i}^s \: v_{i}
	\end{align}
\end{enumerate}

\section{Numerical examples}
\label{sec:results}
\subsection{Smale-Williams solenoid map}
\label{sec:solenoid}
The Smale-Williams solenoid map is a classic 
example of low-dimensional hyperbolic dynamics. 
It is a three-dimensional map given by
\begin{align}
	\varphi^s(u) = \begin{bmatrix}
		s_1 + \dfrac{r-s_1}{4} + \dfrac{\cos(\theta)}{2} \\
		\\
		2 \theta + \dfrac{s_2}{4}\sin(2\pi\theta) \\
		\\
		\dfrac{z}{4} + \dfrac{\sin\theta}{2}
	\end{bmatrix}, 
\end{align}
where $u := [r,\theta, z]^T$ in cylindrical coordinates. 
The attractor is a subset of the solid torus at the 
reference values of $s_1= 1.4$ and $s_2 = 0$. The probability 
distribution on the attractor is not a smooth function but rather 
a generalized function (a distribution) 
of the Sinai-Ruelle-Bowen (SRB) type 
\citep{ruelle-maps,young} that has a density on the unstable manifolds. 
\begin{figure}
%\begin{subfigure}{0.5\textwidth}
\includegraphics[height=7cm,width=\textwidth]{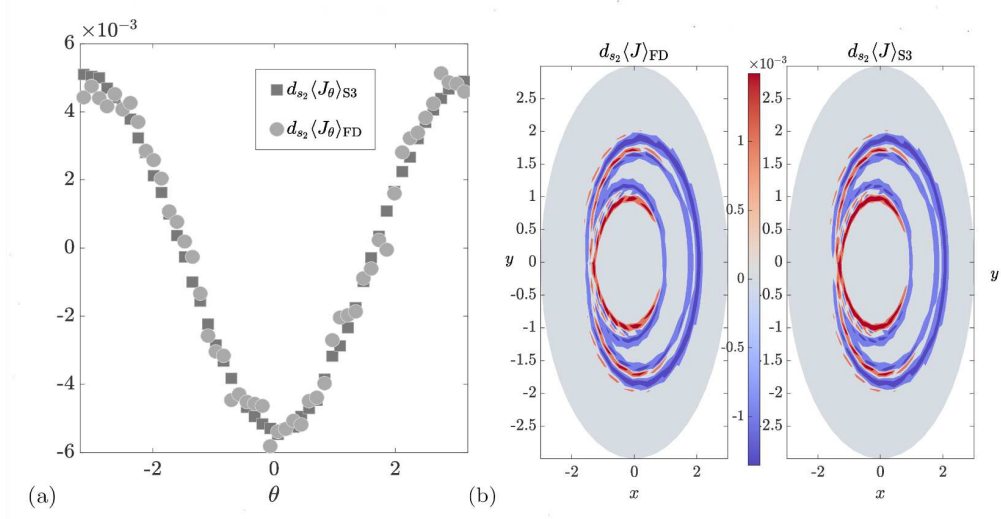}
%\caption{}
%\label{fig:solenoid_dJds2}
%\end{subfigure}
%\begin{subfigure}{0.5\textwidth}
%	\includegraphics[width=\textwidth,height=7cm]{dJtheta_t_CTR}
%	\caption{}
%\label{fig:solenoid_dJtheta_ds2}
%\end{subfigure}

\caption{Comparison of the sensitivities computed with S3 to 
	finite-difference for the solenoid map in Section \ref{sec:solenoid}.
	(a) $J$ is a set of 
	two-variable nodal basis functions along $r$ and $\theta$ axes.
	(b) $J_\theta$ is a set of nodal basis functions 
	along $\theta$ axis. }
\label{fig:solenoid}
\end{figure}

In this map, $r$ and $z$ directions form a basis for 
the stable subspace at each point (and the orthogonal
$\theta$ direction forms a basis for the adjoint unstable subspace).
Applying a perturbation to $s_1$ causes a stable
perturbation, i.e., the unstable contribution is nonzero,
since it affects only the $r$ coordinate. 
On the other hand, perturbing $s_2$ leads to a nonzero
unstable contribution.
A set of nodal basis functions along $r$ and $\theta$ 
is chosen to be the objective function. We use 
a more general S3 algorithm than presented in 
Section \ref{sec:derivation} that is derived under 
the SRB assumption but does not assume the existence of
a density everywhere. In order to validate the S3 computation, 
we compare the sensitivities $(d\langle J \rangle/ds_2)$
with finite-difference results generated using 10 billion
Monte Carlo samples on the attractor. The sensitivities
to the parameter $s_2$ are shown in Figure 1(a). 
In Figure 1(b), the objective function is a set of nodal basis functions
along the $\theta$ direction. From Figures 1(a,b),
we see close agreement between the sensitivities computed with (a more 
general version of) S3 and, finite-difference results, thus validating
both the stable and unstable parts of the S3 algorithm.

%\begin{figure}
%\includegraphics[scale=0.25]{figs/attractor}
%\caption{Plots showing the projection of the solenoid
%attractor on the x-y plane, at different values of 
%the parameters $s_1$ and $s_2$.}
%\label{fig:solenoid_attractor}
%\end{figure}

\subsection{Kuznetsov-Plykin map}
\begin{figure}
%\begin{subfigure}{0.5\textwidth}
%\label{fig:plykin_fd}
\includegraphics[height=6cm,width=\textwidth]{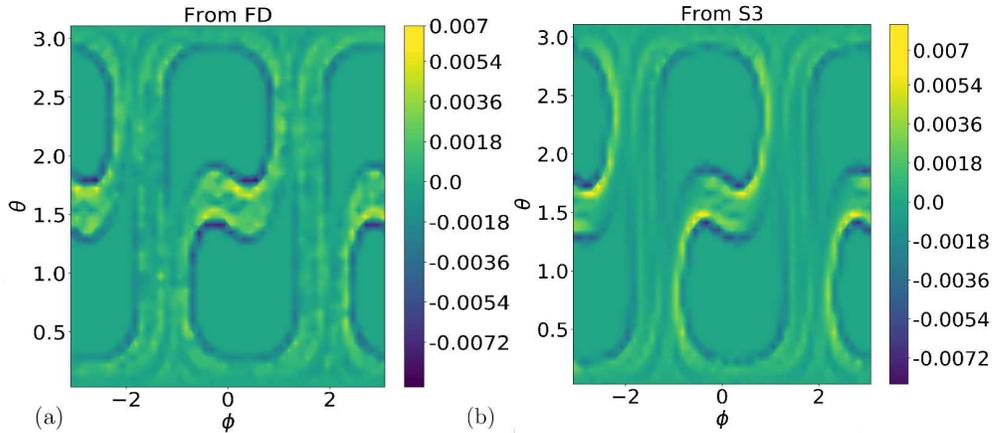}
%\caption{}
%\end{subfigure}
%	\begin{subfigure}{0.5\textwidth}
%	\label{fig:plykin_s3}
%	\includegraphics[height=6cm,width=\textwidth]{dJds_S3}
%	\caption{}
%	\end{subfigure}
\caption{Comparison of the sensitivities of 
	the nodal basis functions 
along the $\theta$ and $\phi$ axes to the parameter $s_2$
	obtained for the Kuznetsov-Plykin attractor using (a) finite difference and (b) the S3 algorithm.}
\label{fig:plykin}
\end{figure}

We consider as a second example the Kuznetsov-Plykin 
map as defined by \cite{kuznetsov}, which describes
a sequence of rotations and translations on the 
surface of the three-dimensional sphere.
The two parameters we choose to vary are $s_1 := \epsilon$ 
and $s_2 := \mu$, 
which are defined by \cite{kuznetsov}. 
The map is given by
\begin{align}
	\varphi_{n+1}^s(u) = f_{-1,-1}\circ f_{1,1}(u),
\end{align}
where $u = [x,y,z]^T$. For the function $f_{\cdot,\cdot}$ and further details regarding the 
hyperbolicity of the system, the reader is 
referred to \cite{kuznetsov}. The probability distribution 
on the attractor again violates the smoothness condition in the derivation but
satisfies the assumption of an existence of a density on the unstable manifolds.
We again use a more general version of the S3 algorithm to compute the 
sensitivities as in the case of the solenoid map in 
Section \ref{sec:solenoid}.
The objective function $J$ is a set of nodal basis functions 
along the $\theta$ and $\phi$
spherical coordinate axes. The finite-difference sensitivities were 
computed with the
central difference around the reference value of $s_2 = 1$ 
by means of 10 billion independent
samples on the attractor. The results from S3 agree well with finite-difference sensitivities 
as shown in Figure \ref{fig:plykin}.

\section{Conclusions}
We have presented the tangent space-split sensitivity algorithm to compute
the sensitivities of statistics to system parameters in chaotic dynamical systems. The algorithm
requires the computation of a basis for the tangent and adjoint unstable subspaces 
along a long trajectory. The stable contribution to the overall sensitivity can be 
efficiently computed by a conventional tangent/adjoint computation just 
as in nonchaotic
systems. The unstable contribution has been derived as an ergodic average that can be evaluated efficiently by using 
solutions to the Koopman tangent 
equation, which has been introduced. The numerical examples described in 
Section \ref{sec:results}
do not satisfy the simplifying assumptions that were made in the derivation.
However, they show close agreement with finite-difference results, suggesting that the ideas used in S3 can be 
extended to more general scenarios.    
%===================================================================
\subsection*{Acknowledgments}
The authors gratefully acknowledge other summer program participants and our
CTR hosts for
many fruitful discussions. We would also like to thank our reviewer 
Dr. Patrick Blonigan for his valuable comments.
%===================================================================
% THE BIBLIOGRAPHY
%=====================================================================
%\bibliographystyle{ctr}
%\bibliography{brief}

\end{document}